\documentclass[prl,superscriptaddress,twocolumn,nofootinbib,showpacs,preprintnumbers,floatfix]{revtex4}

\usepackage{mathrsfs}
\usepackage{amsfonts,amssymb,amsmath}
\usepackage{graphicx,epsfig}
\usepackage{multirow}
\usepackage{verbatim}
\usepackage{slashed}

\declareslashed{}{/}{0}{-0.05}{E}
\declareslashed{}{/}{0}{-0.05}{P}
\declareslashed{}{/}{0}{-0.10}{\Widehat{P}}

\def\fsu5{$\cal{F}$-$SU(5)$}

\def\bfsu5{$\boldsymbol{\mathcal{F}}$-$\boldsymbol{SU(5)}$}

\def\m1half{$M_{1/2}$}
\def\m3half{$M_{3/2}$}
\def\m32{$M_{32}$}

\def\fb{${\rm fb}^{-1}$~}

\def\mt2{$M_{T2}$}
\def\x2{$\chi^2$}
\def\2b{$M_{T2}b$}

\def\bs0{$B_S^0 \rightarrow \mu^+ \mu^-$}

\def\bea{\begin{eqnarray}}
\def\eea{\end{eqnarray}}

\begin{document}

\title{Probing the No-Scale ${\cal F}$-$SU(5)$ One-Parameter Model via Gluino Searches at the LHC2}

\author{Tianjun Li}

\affiliation{CAS Key Laboratory of Theoretical Physics, Institute of Theoretical Physics, 
Chinese Academy of Sciences, Beijing 100190, China}

\affiliation{ School of Physical Sciences, University of Chinese Academy of Sciences, 
No.19A Yuquan Road, Beijing 100049, China}

\author{James A. Maxin}

\affiliation{Department of Chemistry and Physics, Louisiana State University, Shreveport, Louisiana 71115 USA}

\affiliation{Department of Physics and Engineering Physics, The University of Tulsa, Tulsa, OK 74104 USA}

\author{Dimitri V. Nanopoulos}

\affiliation{George P. and Cynthia W. Mitchell Institute for Fundamental Physics and Astronomy, Texas A$\&$M University, College Station, TX 77843, USA}

\affiliation{Astroparticle Physics Group, Houston Advanced Research Center (HARC), Mitchell Campus, Woodlands, TX 77381, USA}

\affiliation{Academy of Athens, Division of Natural Sciences, 28 Panepistimiou Avenue, Athens 10679, Greece}

%%%%%%%%%%%%%%%%%%%%%%%%%%%%%%%%%%%%%%%%%%%%%%%%%%%%%%%%%%%%%%%%%%%%%%%%%%%%

\begin{abstract}

In our recent paper entitled ``The return of the King: No-Scale \fsu5'', we showed that the model space supporting the most favorable phenomenology should have been probed in 2016 at the LHC2, with an even further reach into this region of the model in 2017--18. This ideal realm of the one-parameter version of No-Scale \fsu5 yields a 1.9--2.3~TeV gluino mass at the very same point where the light Higgs boson mass enters its rather narrow experimentally determined range of $m_h = 125.09 \pm 0.24$~GeV. Given the recent results reported at Moriond 2017 for 36 \fb of luminosity collected in 2016 at the 13~TeV LHC2, we now update the status of the No-Scale \fsu5 model space in light of the gluino mass exclusion limits presented by the ATLAS and CMS Collaborations. We illustrate that a resolution could be reached soon as to whether supersymmetry lives in this most critical region of the model space.

\end{abstract}

%%%%%%%%%%%%%%%%%%%%%%%%%%%%%%%%%%%%%%%%%%%%%%%%%%%%%%%%%%%%%%%%%%%%%%%%%%%%


\preprint{ACT-04-17, MI-TH-1755}

\maketitle

%%%%%%%%%%%%%%%%%%%%%%%%%%%%%%%%%%%%%%%%%%%%%%%%%%%%%%%%%%%%%%%%%%%%%%%%%%%%

The ATLAS and CMS Collaborations at the LHC2 recorded 36~\fb of luminosity in 2016 at a center-of-mass energy of 13~TeV. In the search for supersymmetry (SUSY), the LHC2 in 2016 probed gluino ($\widetilde{g}$) masses beyond 2~TeV, with current constraints in the multijet and multiple b-jet search regions residing at 1.92~TeV for ATLAS~\cite{ATLAS:2017vjw} and 1.96~TeV for CMS~\cite{Sirunyan:2017cwe}. It is therefore expected that the LHC2 in 2017 will have at least a reach to a 2.0--2.3~TeV gluino mass. The 13~TeV beams will soon reenergize and begin recording proton-proton collision data again in June 2017, therefore, we use this present short pause in the SUSY search while the LHC2 prepares to launch its 2017 run to evaluate the model space of one prominent supersymmetric Grand Unified Theory (GUT) model against the full compilation of 2016 data statistics.

The supersymmetric GUT model No-Scale flipped $SU(5)$ with extra vector-like $flippon$ multiplets, dubbed \fsu5 for short, was shown in ``The return of the King: No-Scale \fsu5''~\cite{Li:2016bww} to possess the curious characteristic of generating an $m_{\widetilde{g}} = 1.9-2.3$~TeV gluino mass at the very point in the model space where the light Higgs boson mass enters into its experimentally viable $1\sigma$ range of $m_h = 125.09 \pm 0.24$~GeV~\cite{:2012gk,:2012gu}. In GUTs with gravity mediated SUSY breaking, otherwise known as the Minimal Supergravity (mSUGRA) models, the SUSY breaking soft terms can be represented by four universal terms: gaugino mass $M_{1/2}$, scalar mass $M_0$, trilinear soft term $A_0$, and the low-energy ratio of the Higgs vacuum expectation values (VEVs) tan$\beta$, in addition to the sign of the Higgs bilinear mass term $\mu$. The most elemental No-Scale scenario requires $M_0 = A_0 = B_{\mu} = 0$ at the ultimate high-energy unification scale $M_{\cal F}$, where $B_{\mu}$ is the SUSY breaking soft term for the $\mu H_d H_u$ term in the superpotential. All the SUSY breaking soft terms therefore evolve down to low-energy from the single non-zero parameter $M_{1/2}$ at the $M_{\cal F}$ scale, hence, a true one-parameter model. Consequently, the entire SUSY spectrum will be proportional to $M_{1/2}$ at the leading order, yielding an invariance of the bulk ``internal'' physical properties under an overall rescaling. The complete SUSY spectrum thus maintains a piecewise smooth linear functional dependence upon $M_{1/2}$. A gluino mass of $m_{\widetilde{g}} = 1.9-2.3$~TeV will therefore correlate to only a single specific confined range of light Higgs boson masses, which in the case of No-Scale \fsu5 happens to be its observed value. In addition to reproducing the observed light Higgs boson mass, the $m_{\widetilde{g}} = 1.9-2.3$~TeV gluino mass range under analysis here further produces the 9-year WMAP~\cite{Hinshaw:2012aka} and Planck 2015~\cite{Ade:2015xua} precision relic density $1\sigma$ measurements, while also conforming with empirical data for rare-decay processes, direct dark matter detection cross-sections, and proton decay lifetimes~\cite{Li:2016bww}.

\begin{figure*}[htp]
        \centering
        \includegraphics[width=1.0\textwidth]{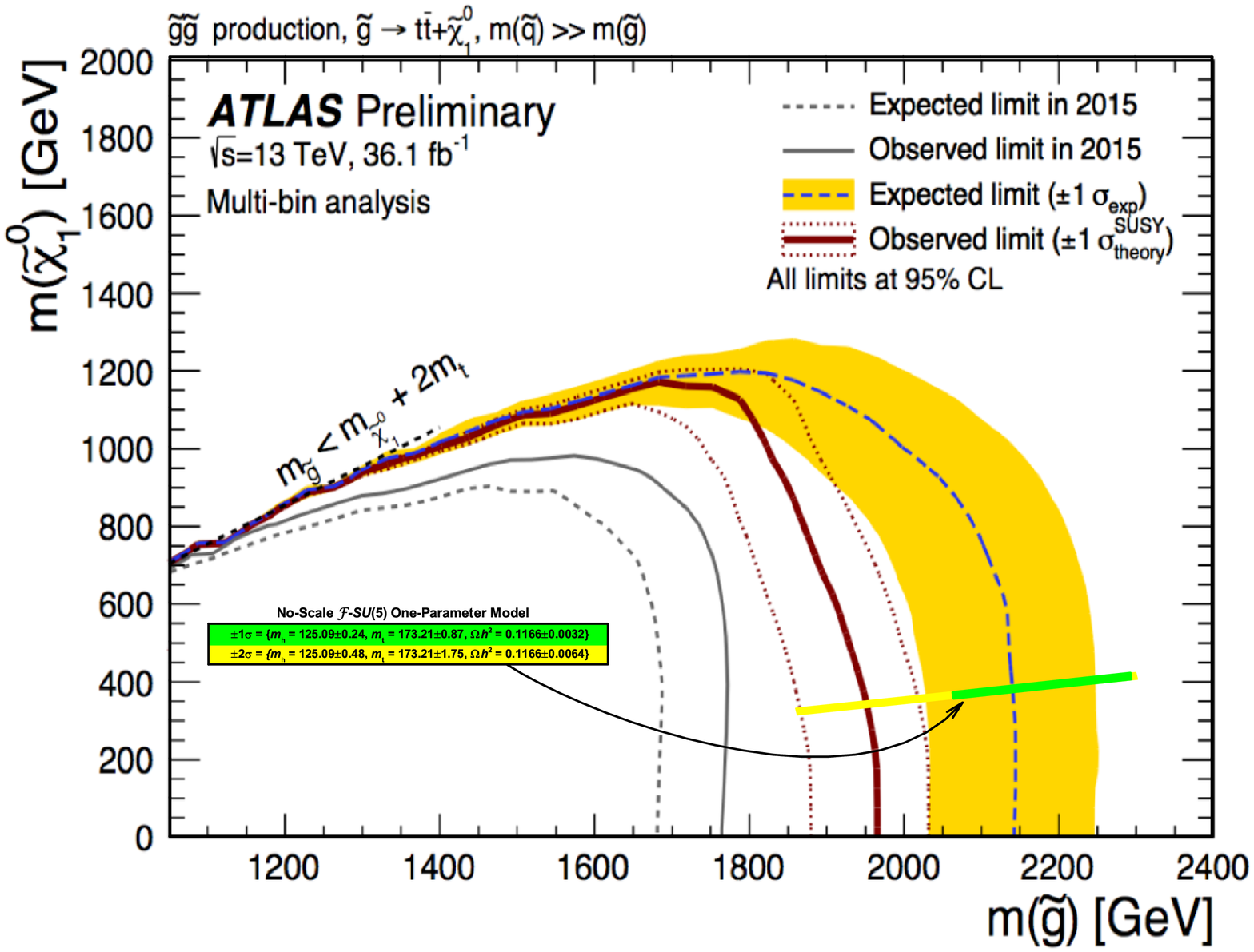}
        \caption{The ATLAS 95\% CL expected and observed exclusion limits of Ref.~\cite{ATLAS:2017vjw} with the linear model space of No-Scale \fsu5 superimposed. The green strip of model space is consistent with the $\pm 1 \sigma$ uncertainties on the light Higgs boson mass, top quark mass, and cold dark matter relic density. The yellow strip is that region of the model space satisfying the $\pm 2 \sigma$ uncertainties on these same experiments.}
        \label{fig:atlas}
\end{figure*}

The accompanying vector-like multiplets, which we denote with the characterful term $flippons$ when the multiplets are explicitly implemented within No-Scale flipped $SU(5)$, induce a few distinctive phenomenological signatures. As is known with No-Scale Supergravity, the light stau will be the Lightest Supersymmetric Particle (LSP) if the supersymmetry breaking scale is the traditional GUT scale around $2.0\times 10^{16}$~GeV. Interestingly, the lightest neutralino with a Bino dominant component becomes the LSP in our model since the supersymmetry breaking scale is around the string scale, and then we can obtain the correct dark matter density via the light stau and LSP neutralino coannihilation~\cite{Li:2010ws, Li:2010mi}. Moreover, we run the Renormalization Group Equations (RGEs) from the string scale to the $SU(3)_C\times SU(2)_L$ unification scale $M_{32}$, {\it i.e.}, around the traditional GUT scale. Therefore, at the traditional GUT scale, we shall have non-Universal supersymmetry breaking soft terms, unlike mSUGRA or the Constrained Minimal Supersymmetric Standard Model (CMSSM)~\cite{Li:2010ws, Li:2010mi}. In addition, the one-loop beta function for the $SU(3)_C$ gauge symmetry receives a null value after inclusion of the flippons, {\it i.e.}, $b_3 = 0$, and thus the gaugino mass $M_3$ mirrors the vanishing $b_3$ via flat RGE evolution, as the conventional low-energy logarithmic mass enhancement is suppressed. The outcome is a scaled-down gluino mass reduced below the first/second generation squarks, bottom squarks, and heavy stop, preserving a positive mass delta with only the light stop $\widetilde{t}_1$. Given this rather unique SUSY mass spectrum of $M(\widetilde{t}_1) < M(\widetilde{g}) < M(\widetilde{q})$, where $\widetilde{q} = \{ \widetilde{u}, \widetilde{d}, \widetilde{c}, \widetilde{s}, \widetilde{b}_1, \widetilde{b}_2, \widetilde{t}_2 \}$, a compulsory gluino decay through an on-shell light stop of $\widetilde{g} \to \widetilde{t}_1 t \to t \overline{t} + \widetilde{\chi}_1^0$ is imposed, triggering a hard 4-top multijet event via pair-produced gluinos and/or first/second generation squarks per $\widetilde{q} \to q \widetilde{g}$. The 4-top 100\% branching fraction ensures a prolific accumulation of events with large numbers of hadronic jets, including tagged b-jets. This No-Scale \fsu5 characteristic SUSY event topology compels a search region at the LHC2 inclusive of final states with multijet events and multiple b-jets. For a more detailed background of the No-Scale \fsu5 model, please see Refs.~\cite{Maxin:2011hy,Li:2011ab,Li:2013naa,Leggett:2014hha,Li:2016bww} and references therein.

\begin{figure*}[htp]
        \centering
        \includegraphics[width=1.0\textwidth, trim={0 0 0.93cm 0},clip=true]{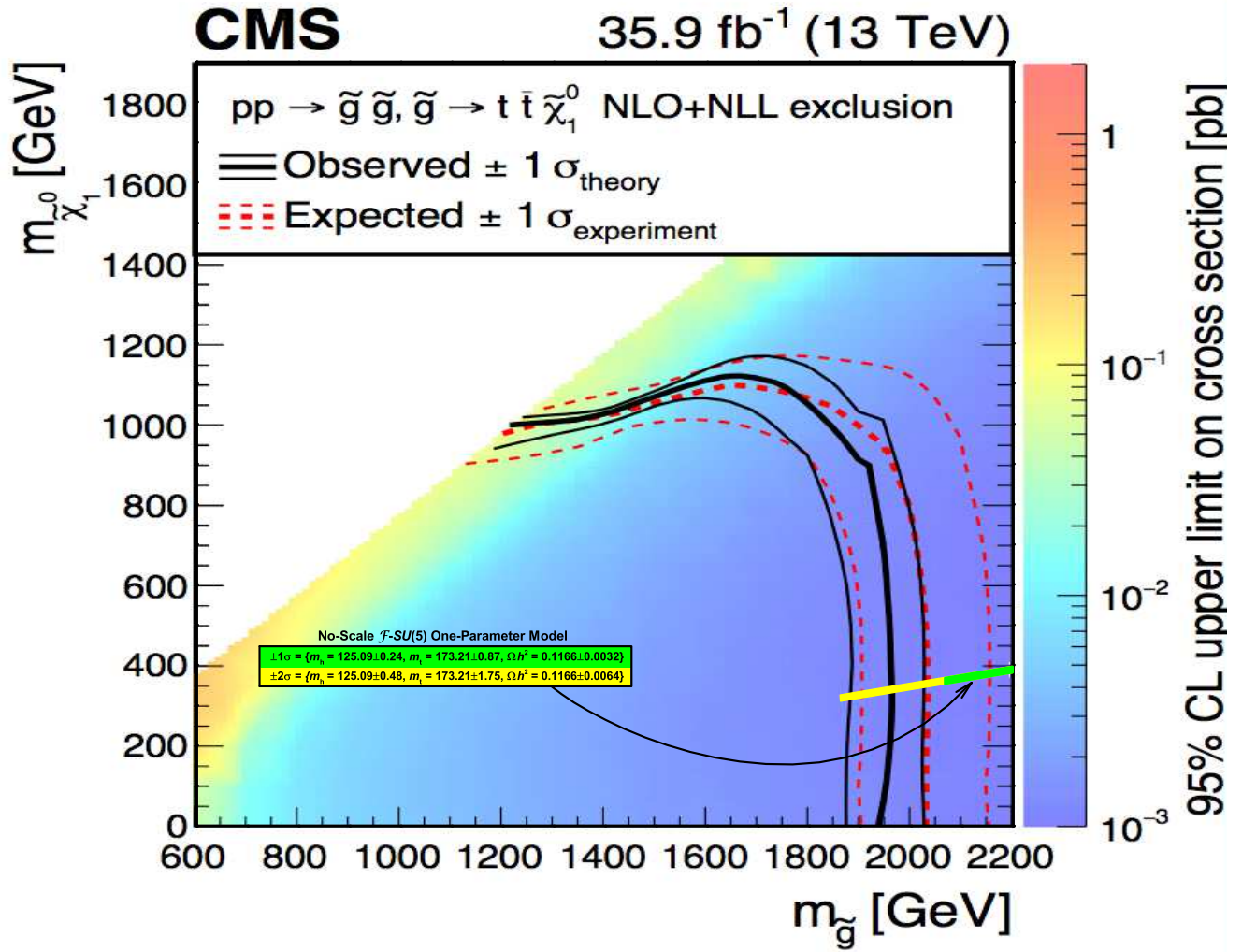}
        \caption{The CMS 95\% CL expected and observed exclusion limits of Ref.~\cite{Sirunyan:2017cwe} with the linear model space of No-Scale \fsu5 superimposed. The green strip of model space is consistent with the $\pm 1 \sigma$ uncertainties on the light Higgs boson mass, top quark mass, and cold dark matter relic density. The yellow strip is that region of the model space satisfying the $\pm 2 \sigma$ uncertainties on these same experiments.}
        \label{fig:cms}
\end{figure*}

The analysis of the multijet and multiple b-jet SUSY search regions for the 36~\fb 2016 run at the LHC2 have been completed by ATLAS~\cite{ATLAS:2017vjw} and CMS~\cite{Sirunyan:2017cwe}. Both studies focus on gluino pair-production leading to a 4-top signature by means of $\widetilde{g} \to t \overline{t} + \widetilde{\chi}_1^0$, as annotated in FIG.~\ref{fig:atlas} and FIG.~\ref{fig:cms}. The ATLAS 95\% CL expected and observed exclusion limits of Ref.~\cite{ATLAS:2017vjw} are reproduced here in FIG.~\ref{fig:atlas}, showing the $\pm1 \sigma$ boundaries enclosing the expected and observed limits, where ATLAS refers to this search region as the Gtt simplified model. The CMS 95\% CL expected and observed exclusion limits of Ref.~\cite{Sirunyan:2017cwe} are also reproduced here in FIG.~\ref{fig:cms}, displaying the $\pm1 \sigma$ uncertainties surrounding the expected and observed limits, where CMS refers to this search region as the T1tttt simplified model. We superimpose on the ATLAS and CMS exclusion plots in FIGs.~\ref{fig:atlas}--\ref{fig:cms} the linear model space of No-Scale \fsu5 as described in the prior paragraphs. We distinguish between the regions of the model space that satisfy the $\pm1 \sigma$ and $\pm2 \sigma$ experimental uncertainties for a set of experiments that comprise the light Higgs boson mass, top quark mass, and total cold dark matter relic density. The light Higgs boson mass central value and $\pm1 \sigma$ uncertainty is $m_h = 125.09 \pm 0.24$~GeV~\cite{:2012gk,:2012gu}, where for the $\pm2 \sigma$ limits we double the $1 \sigma$ experimental uncertainty and add/subtract from the observed central value. For the top quark mass, we apply the PDG 2016 $1 \sigma$ world average of $m_t = 173.21 \pm 0.874$~GeV~\cite{Olive:2016xmw}, also doubling the $1 \sigma$ uncertainty for the $2 \sigma$ limits. The relic density constraint we employ exploits both the 9-year WMAP~\cite{Hinshaw:2012aka} and Planck 2015~\cite{Ade:2015xua} measurements, using the WMAP9 $1 \sigma$ lower limit as our minimum value and the Planck 2015 $1 \sigma$ upper limit as our maximum value. This provides a relic density constraint of $\Omega h^2 = 0.1166 \pm 0.0032$. In FIGs.~\ref{fig:atlas}--\ref{fig:cms} the region of the model space satisfying the $\pm 1 \sigma$ experimental constraints $\{m_h = 125.09 \pm 0.24,~m_t = 173.21 \pm 0.874,~\Omega h^2 = 0.1166 \pm 0.0032 \}$ is highlighted in green, whereas all the points in the model space consistent with the $\pm 2 \sigma$ experimental constraints $\{m_h = 125.09 \pm 0.48,~m_t = 173.21 \pm 1.75,~\Omega h^2 = 0.1166 \pm 0.0064 \}$ are accentuated in yellow, which of course will further encompass the green $\pm 1 \sigma$ strip as well.

The upper $1 \sigma$ limit on the ATLAS and CMS exclusion limits delineated in FIGs.~\ref{fig:atlas}--\ref{fig:cms}  lands at about 2030~GeV. Evidently, according to Ref.~\cite{ATLAS:2017vjw} the observed limits were significantly weaker than expected attributable to mild excesses observed in the relevant signal regions. The looser than anticipated limits are clearly exhibited in FIGs.~\ref{fig:atlas}--\ref{fig:cms}, certainly right on the cusp of that precise region of No-Scale \fsu5 compatible with the $\pm 1 \sigma$ experimental uncertainties on the light Higgs boson mass, top quark mass, and relic density, as this region of No-Scale \fsu5 was expected to be mostly excluded at $1 \sigma$ in the 36 \fb 2016 run but could not due to the mild excesses noted above. Consequently, this realm of No-Scale \fsu5 satisfying the $\pm 1 \sigma$ empirical uncertainties remains viable and it is reasonable to conclude the region was probed in 2016 and surely the LHC2 will reach deeply into this region in 2017--18. We thus can confidently project that the presence of SUSY in this rather phenomenologically favorable dominion of No-Scale \fsu5 should be adjudicated in the forthcoming couple of years, either to the delight or dismay of SUSY enthusiasts near and far.

%%%%%%%%%%%%%%%%%%%%%%%%%%%%%%%%%%%%%%%%%%%%%%%%%%%%%%%%%%%%%%%%%%%%%%%%%%%%

\section{Acknowledgments}

The primary numerical results for this project were computed at the Tandy Supercomputing Center, using dedicated resources provided by The University of Tulsa.  Other supporting data for the theoretical background have been obtained via the HPC Cluster of ITP-CAS. This research was supported in part by the Projects 11475238 and 11647601 supported by the National Natural Science Foundation of China and Key Research Program of Frontier Sciences CAS (TL), and by DOE grant DE-FG02-13ER42020 (DVN).

%%%%%%%%%%%%%%%%%%%%%%%%%%%%%%%%%%%%%%%%%%%%%%%%%%%%%%%%%%%%%%%%%%%%%%%%%%%%

\bibliography{bibliography}

\end{document}